\documentclass[aps,prb,preprint,showpacs,superscriptaddress,showkeys,12p]{revtex4-1}
\usepackage{amssymb,amsmath,amsbsy,graphicx,hyperref,xcolor,csquotes,braket,epstopdf}
\usepackage{subcaption}

\begin{document}
	
\title{Helical edge states in silicene and germanene nanorings in perpendicular magnetic field}

\author{Du\v{s}an Z. Jakovljevi\'c}\email{jakovljevic@etf.bg.ac.rs}
\affiliation{School of Electrical Engineering, University of Belgrade, P.O. Box 3554, 11120 Belgrade, Serbia}
\author{Marko M. Gruji\'c}%\email{marko.grujic@etf.bg.ac.rs}
\affiliation{School of Electrical Engineering, University of Belgrade, P.O. Box 3554, 11120 Belgrade, Serbia}
\author{Milan \v{Z}. Tadi\'c}%\email{tadic@etf.bg.ac.rs}
\affiliation{School of Electrical Engineering, University of Belgrade, P.O. Box 3554, 11120 Belgrade, Serbia}
\author{Fran\c cois M. Peeters}%\email{francois.peeters@uantwerpen.be}
\affiliation{Department of Physics, University of Antwerp, Groenenborgerlaan 171, B-2020 Antwerp, Belgium}

\begin{abstract}
Due to nonzero intrinsic spin-orbit interaction in buckled honeycomb crystal structures, silicene and germanene exhibit  interesting topological properties, and are therefore candidates for the realization of the quantum spin Hall effect. We employ the Kane-Mele model to investigate the electron states in hexagonal silicene and germanene nanorings having either zigzag or armchair edges in the presence of a perpendicular magnetic field. We present results for the energy spectra as function of magnetic field, the electron density of the spin-up and spin-down states in the ring plane, and the calculation of the probability current density. The quantum spin Hall phase is found at the edges between the nontrivial topological phase in silicene and germanene and vacuum. We demonstrate that the helical edge states in zigzag silicene and germanene nanorings can be qualitatively well understood by means of classical magnetic moments. However, this is not the case for comparable-sized armchair nanorings, where the eigenfunctions spread throughout the ring. Finally, we note that the energy spectra of silicene and germanene nanorings are similar and that the differences between the two are mainly related to the difference in magnitude of the spin-orbit coupling.
\end{abstract}

\maketitle

\section{Introduction}\label{I}

Silicene and germanene are monolayers of silicon and germanium whose atoms are arranged in a honeycomb lattice. But in both cases the lattice is buckled, which along with the relative heaviness of the atoms leads to a nonzero intrinsic spin-orbit coupling (SOC).\cite{cahangirov09,liu11prl} The SOC opens up a gap at the Dirac points and the low-energy electron states behave as massive Dirac quasi-particles.

Due to the nonzero intrinsic SOC silicene and germanene are two-dimensional topological insulators.\cite{hasan10,qi11} Their energy gap is insulating, yet if there is a boundary separating them from a topologically different phase, which has different $\mathbb{Z}_2$ invariant,\cite{kaneZ2} vacuum for example, gapless states emerge at the sample edges.\cite{kaneQSH} These states are counterpropagating and have opposite spins, which implies they are helical in nature, and thus favorable for the formation of the quantum spin Hall (QSH) state. This can be understood in terms of the {\it bulk-edge correspondence}, which states that connecting two topologically different insulators involves a phase transition that leads to a vanishing energy gap. This correspondence has been previously evoked to study zero-dimensional nanodisks.\cite{kikutake13} Strictly speaking, the topological invariant is ill-defined for a nanostructure, but if the system is sufficiently large the helical states emerge at the edges. The protected edge states polarized in valley-spin pairs are shown to appear  in silicene dots and antidots in the presence of a perpendicular magnetic field.\cite{rakyta15} And loops delineating two topologically distinct insulating domains were demonstrated to have chiral properties.\cite{grujic15} Also, the topological properties of the confined honeycomb lattice in the form of a rectangular plaquette having both zigzag and armchair edges were studied in the presence of a perpendicular magnetic field and taking into account intrinsic SOC.\cite{ostahie14} The results obtained in Ref.~\onlinecite{ostahie14} motivated us to investigate how helical states appear at nanoring edges, and how they interplay with an external magnetic field.

In this paper, hexagonal nanorings made of silicene or germanene are considered. A perpendicular magnetic field is applied, and we investigate how the results depend on the type of ring edges, i.e. either zigzag or armchair. Length of the outer and inner zigzag edges is expressed as the number of six-atom rings along the sides of the respective hexagons, $N_E$ and $N_I$, respectively. For the armchair edge $N_E$ and $N_I$ denote the numbers of the outermost six-atom rings along the nanoring sides. We explain the obtained data qualitatively in terms of classical magnetic moments with the aim to show that the spin-up and spin-down electrons correspond to helical edge states in nanorings. Also, we study the dependence of the energy levels on the type of edges and contrast the results for silicene rings with those for germanene rings. The found differences can be related to the magnitude of spin-orbit coupling.

\section{The model} \label{II}

\subsection{The Hamiltonian} \label{A}

To study silicene and germanene nanorings we used the Kane-Mele model which is suitable for honeycomb lattices, and we took into account the presence of a perpendicular magnetic field through the Peierels phase. It is the $\pi$-band tight-binding model (TBM) which takes into account SOC,
\begin{equation}
H = -t \sum_{\langle n,m\rangle} e^{i\varphi_{nm}} c^{\dagger}_nc_m + i\frac{s\Delta_{SO}}{3\sqrt{3}} \sum_{\langle\langle n,m \rangle\rangle} \nu_{nm} e^{i\varphi_{nm}} c^{\dagger}_nc_m.
\end{equation}
Here $s=1$ labels the spin up and $s=-1$ the spin down. The first term refers to the nearest-neighbour (NN) hopping between the $p_z$ orbitals, with $t=1.6$ eV and $t=1.3$ eV for silicene and germanene, respectively.\cite{liu11} The second term takes into account SOC, where $\Delta_{SO}$ for silicene and germanene is given by 3.9 meV and 46.3 meV, respectively.\cite{liu11} The parameter $\nu_{nm}$ refers to the direction of the next nearest neighbour (NNN) hopping, and amounts to $+1$ and $-1$ for an electron hopping from a site $m$ to a site $n$, making a right and a left turn, respectively. The Peierls term $\varphi_{nm}=-\frac{e}{\hbar} \int_{\mathbf{r}_m}^{\mathbf{r}_n} \mathbf{A} \cdot \mathrm{d}\mathbf{l}$ is the phase that the electron acquires while travelling in the presence of a perpendicular magnetic field $\mathbf{B}=B\mathbf{e}_z$. The value of the lattice constant $a$, which is the NNN distance, amounts to $a=3.86$ \AA \, in silicene and $a=4.02$ \AA \, in germanene.\cite{liu11}

\subsection{Probability current density}\label{B}

To supplement our analysis, probability current density (PCD), $\bf{j}$, is computed using the model adapted from Refs. [\onlinecite{dacosta12}] and  [\onlinecite{dacosta14}]. The $x$ and $y$ components of $\bf{j}$ at site $n$ are given by
\begin{equation}
j_{x(y)}(n)=\sum_{m} \frac{2d_{x(y)}}{\hbar} \, \mathrm{Im}[C_n C_m^{*} H_{mn}],
\label{PDC}
\end{equation}
where $C_{n}$ is the expansion coefficient related to the atomic site $n$, the index $m$ runs through all three NN and six NNN lattice positions, and $d_x = x_m - x_n$ and $d_y = y_m - y_n$ are the distances between the sites $n$ and $m$ along the $x$ and $y$ direction, respectively. Note that the definition of PCD in Eq.~(\ref{PDC}) satisfies the continuity equation. Also, it is straightforward to deduce that if the electron hops from site $n$ to site $m$ in the $x(y)$ direction, the $x(y)$ component of $j$ is positive.

%DOVDE

\section{Silicene nanoring with zigzag edges}\label{III}

\begin{figure}
	\centering
	\includegraphics[width=15cm]{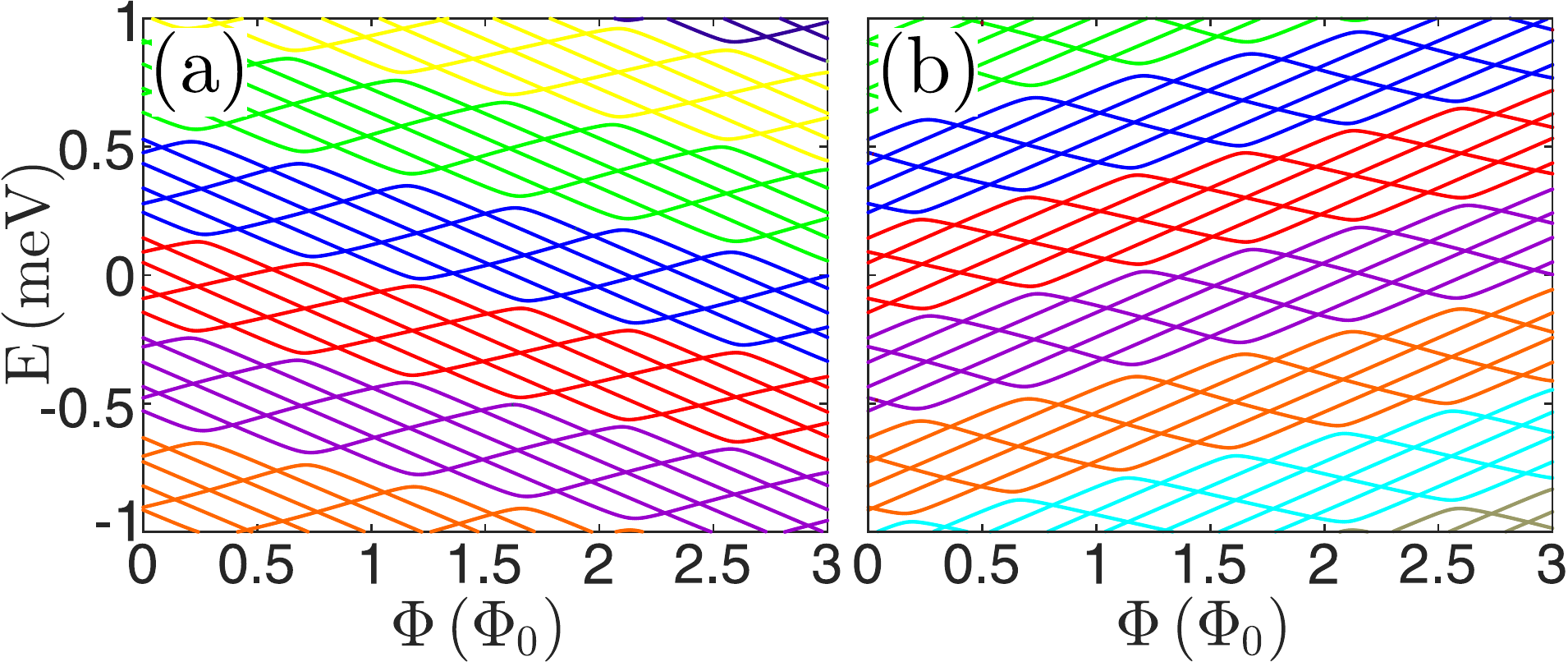}
	\caption{The energy levels of a silicene nanoring having zigzag edges as function of magnetic flux for: (a)  spin-up, and (b) spin-down electron states. Magnetic flux through the ring opening is in units of the flux quantum $\Phi_0$. Different subbands are denoted by different colors.}
	\label{fig1}
\end{figure}

We first computed the energy levels in silicene nanorings with zigzag edges. For computational purposes, the outer and inner edge are taken to be $N_E = 60$ and $N_I=30$ long, which is a reasonable choice regarding previously analyzed graphene nanorings. The energies of the spin-up and spin-down states are shown in Fig. \ref{fig1} as function of magnetic flux threading the ring opening. Magnetic flux through the ring opening, $\Phi$, is expressed in the number of flux quanta $\Phi_0=h/e$, where $\Phi = \Phi_0$ corresponds to $B \approx 12$ T. Note that we only depicted the energy levels inside the SO gap ($-\Delta_{SO} \leqslant E \leqslant \Delta_{SO}$) surrounding the Fermi energy $E_F=0$ eV.  The spin degeneracy is obviously lifted in the magnetic field because of the nonzero SOC. The two spectra are mirror imaged with respect to the Fermi energy, and the levels are arranged in groups (subbands) of six due to the ring's sixfold rotational symmetry. The average energies of the spin-up (spin-down) subbands decrease (increase). Note that the intrinsic SOC is responsible for opening a gap, in which edge states can appear (see Fig.~\ref{fig1}), whereas $\Delta_{SO} = 0$ in graphene, thus no similar electronic structure emerges in graphene nanorings. But in the range $|E|>\Delta_{SO}$ the spectra of silicene and graphene nanorings appear to be similar (see  Refs.~\onlinecite{dacosta14} and \onlinecite{bahamon09}), and therefore will not be considered here, i.e. we focus on the energy levels inside the SO gap.
%In magnetic field the spectrum depends on the ring size and the SOC strength. The states can be classified in three classes  showing large mutual differences: (i) the states whose energy decreases with magnetic flux, (ii) the states with energy increasing with flux, and (iii) the states at points of anticrossings. The distributions of the eigenfunctions in the ring plane typical for the three states and for the two spins are displayed and discussed. We also include the plots of the probability current density for the three states.

\begin{figure}
\centering
\includegraphics[width=9cm]{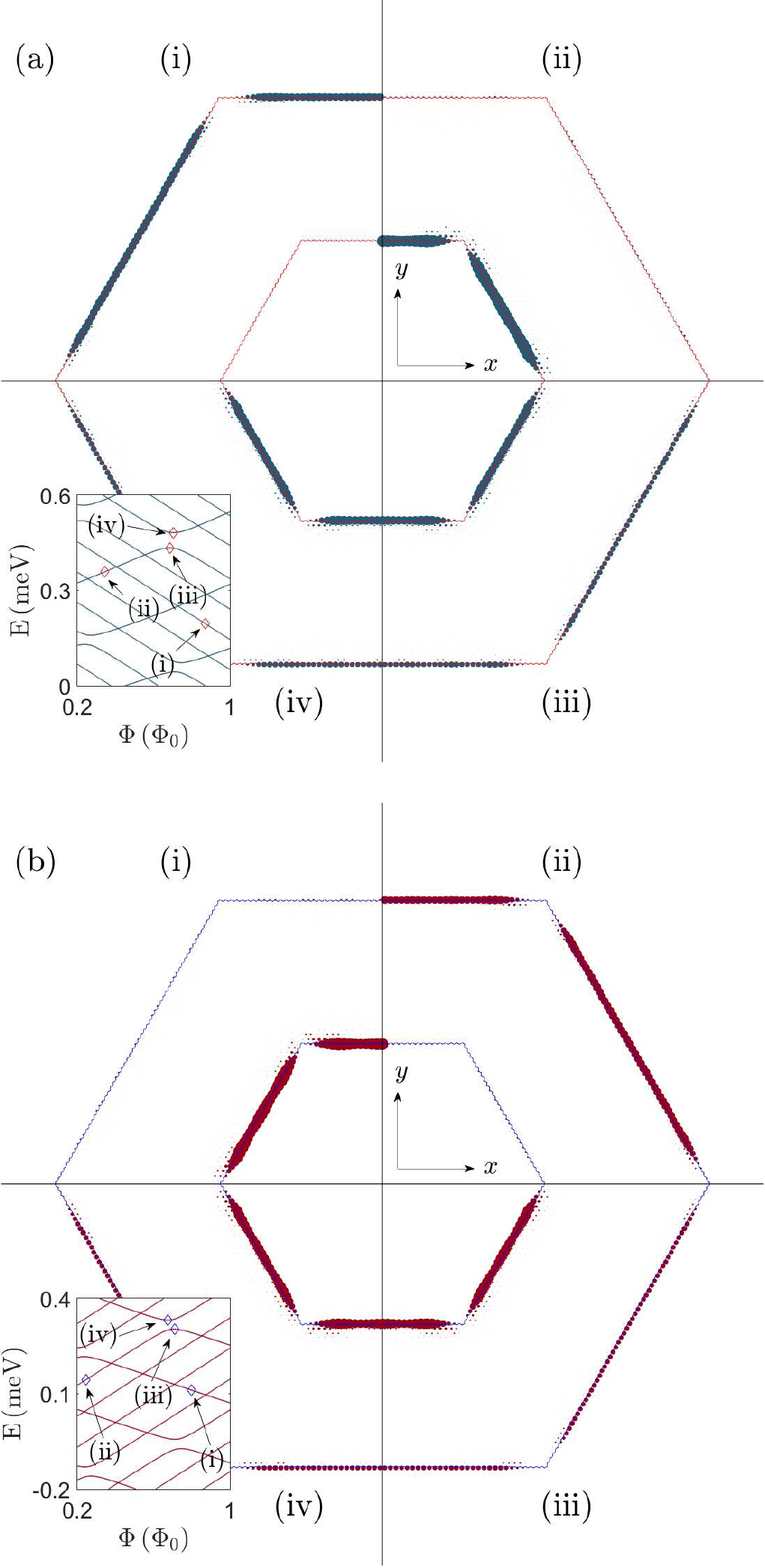}
\caption{(a) The electron densities for the spin-up states (i)-(iv) are displayed in different sections of the nanoring. The inset: a zoom of the spectrum of the spin-up electron states shown in Fig. \ref{fig1}(a) with the states (i)-(iv) indicated by arrows. (b) The same as (a), but for the spin-down states.}
\label{fig2}
\end{figure}

Ref.~\onlinecite{bahamon09} showed that graphene nanorings have a state localized at the nanoring outer edge whose energy decreases with $B$, whereas the energy of a state localized at the ring inner edge increases with $B$. An anticrossing between those levels takes place when the edges of the nanoring become coupled. From the plot of the electron density for the chosen states in Fig.~\ref{fig2}(a), we found that the same occurs for the spin-up states in silicene nanorings. To illustrate this, a detail of the spectrum shown in Fig.~\ref{fig1}(a) is displayed in the inset of Fig.~\ref{fig2}. The electron densities of four states denoted by (i)-(iv) are shown to illustrate the formation of the helical edge states. Here, (i) is a state whose energy decreases with flux, the state (ii) has energy increasing with flux, and (iii) and (iv) are states which anticross with each other. The inset in Fig.~\ref{fig2}(a) shows that the states of the same subband exhibit crossings whereas anticrossings occur between states belonging to different subbands. Since the TBM-determined states are not labelled by any quantum number, except for the principal one, it is not evident to classify crossings or anticrossings by specific quantum numbers.

The nanorings with zigzag edges in Fig.~\ref{fig2} are divided into four sections, each displaying a particular electron density distribution. The distribution in the whole nanoring plane can be  constructed from those segments and using the six-fold symmetry of the nanoring. The plotting convention for the density reads: the radius of a circle centered at a certain atomic site $n$ corresponds to the modulus squared of the amplitude expansion coefficient, $\vert C_n\vert^2$, and is  proportional to the local probability density. The electron density of the states (i) and (ii) shown in the upper sections of Fig.~\ref{fig2}(a) are clearly localized at the ring outer and inned edge, respectively. The distributions of the states (iii) and (iv), shown in the lower sections of Fig. \ref{fig2}(a), are localized at both edges, which is a signature that the edges are coupled in these states. Acctually, both states have very similar electron distribution, as in the case of bonding and anti-bonding states in e. g. a coupled well system.

%\begin{figure}
	%\centering
	%\begin{subfigure}[b]{0.6\textwidth}
		%\includegraphics[width=9cm]{fig2_tif.eps}
		%\caption{}
	%\end{subfigure}
	%\qquad %add desired spacing between images, e. g. ~, \quad, \qquad, \hfill etc.
	%(or a blank line to force the subfigure onto a new line)
	%\begin{subfigure}[b]{0.6\textwidth}
		%\includegraphics[width=9cm]{Figure2b_samo_ivica.eps}
		%\caption{}
	%\end{subfigure}
	%\caption{(a) The electron densities for the spin-up states (i)-(iv) are displayed in different sections of the nanoring. The inset: a zoom of the spectrum of the spin-up electron states shown in Fig. \ref{fig1}(a) with the states (i)-(iv) indicated by arrows. (b) The same as (a), but for the spin-down states.}
	%\label{fig2}
%\end{figure}

%\begin{figure}[t!]
	%\centering
	%\includegraphics[width=12cm]{Figure2.eps}
	%\caption{The electron densities for the states (i)-(iv) are displayed in different sections of the nanoring. The inset: a zoom of the spectrum of the spin-up electron states shown in Fig. \ref{fig1}(a) with the states (i)-(iv) indicated by arrows.}
	%\label{fig2}
%\end{figure}

The electron distribution of the given spin state shown in Fig.~\ref{fig2}(a) can be interpreted within the model of a classical magnetic dipole in a magnetic field. The dipole's energy is $E=-\boldsymbol{\mu} \cdot \mathbf{B}$, where $\boldsymbol{\mu}$ is the dipole's magnetic moment. If $\partial E/ \partial B < 0$ the magnetic moment and the magnetic field are parallel, and the persistent current circulates counterclockwise around the dipole. For the opposite case, $\partial E/ \partial B > 0$, the magnetic moment and the magnetic field are antiparallel, implying that the current flows clockwise. We might therefore conclude that the state (i) localized at the outer edge depicted in Fig.~\ref{fig2}(a) carries the persistent current $I^E_{\uparrow}$ in the counterclockwise direction, whereas the state (ii) localized at the inner edge carries the persistent current $I^I_{\uparrow}$ in the clockwise direction. And the states (iii) and (iv) exhibit hybridization between the two edge states, which is manifested in the persistent currents circulating around both nanoring edges in opposite directions. Also, the currents in the states (iii) and (iv) circulate in the same direction at the same edge.

%\begin{figure}[t!]
	%\centering
	%\includegraphics[width=12cm]{Figure2b.eps}
	%\caption{The same as Fig.~\ref{fig2}, but for the spin-down states.}
	%\label{fig3}
%\end{figure}

The densities of the selected spin-down states in the ring plane are shown in Fig. \ref{fig2}(b), with their eigenenergies indicated by the arrows in the inset. The spin-down state (i) exhibits the persistent current $I_{\downarrow}^I$, which circulates counterclockwise around the inner edge (see upper left segment of Fig.~2(b)). On the other hand, the state (ii) carries the current $I_{\downarrow}^E$ which flows clockwise around the outer edge (see upper right segment). And similar to the spin-up case the conductive channels of the states (iii) and (iv) are localized at both nanoring edges with persistent currents at different edges in each state flowing opposite to each other, which is shown in lower segments of Fig.~\ref{fig2}(b).

\begin{figure}[t!]
	\centering
	\includegraphics[width=15cm]{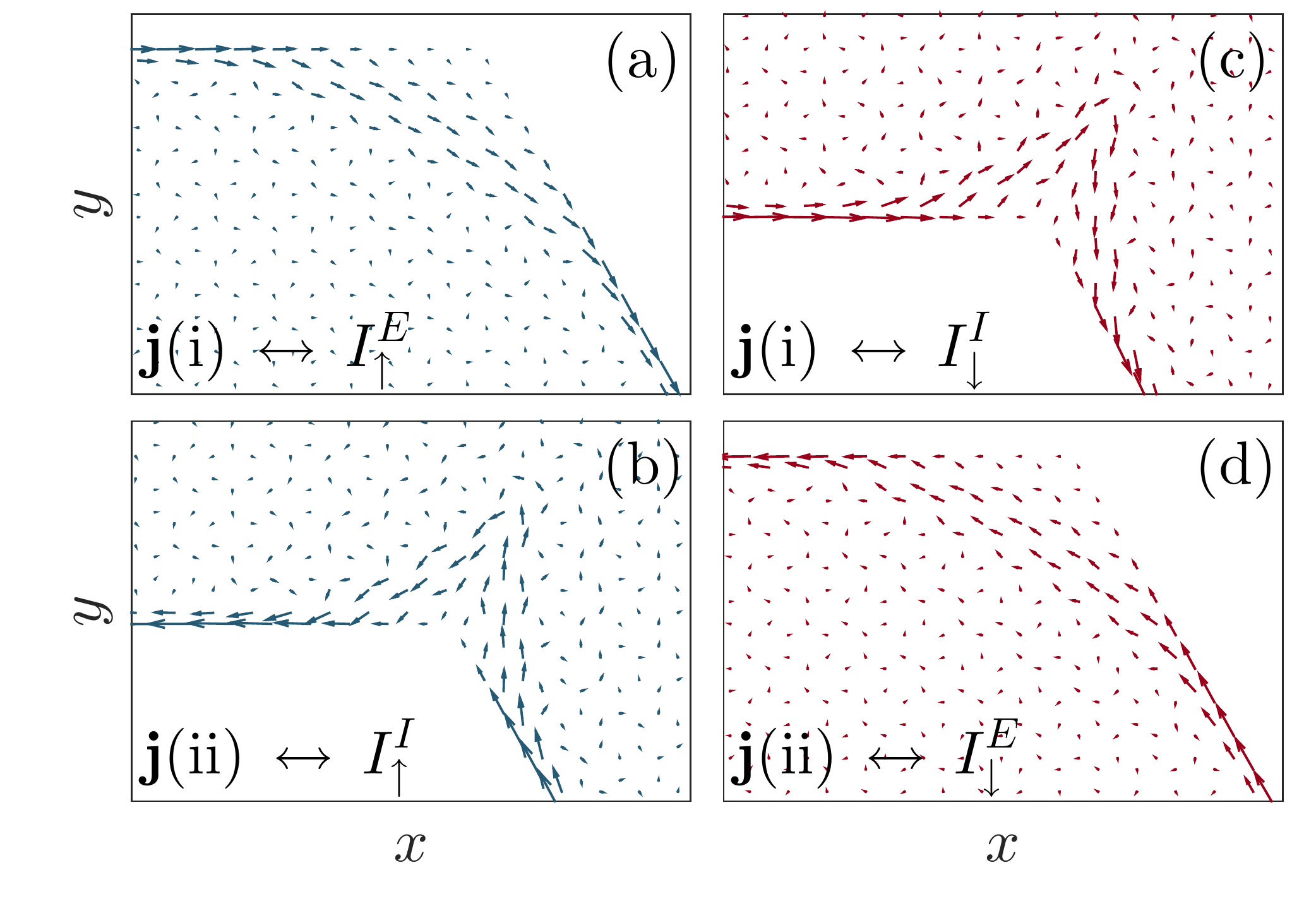}
	\caption{The probability current density distribution around the ring corner for: (a) the persistent current $I_\uparrow^E$ (state (i) in Fig. \ref{fig2}(a)), (b) the persistent current $I_\uparrow^I$ (state (ii) in Fig. \ref{fig2}(a)), (c) the persistent current $I_\downarrow^I$ (state (i) in Fig. \ref{fig2}(b)) and (d) the persistent current $I_\downarrow^E$ (state (ii) in Fig. \ref{fig2}(b)).}
	\label{fig3}
\end{figure}

The classical view in the directions and the localizations of the persistent currents based on classical magnetic dipoles interacting with magnetic field is supported by the PCD distributions as shown in Fig.~\ref{fig3}. Here, note that the PCD determined from Eq.~(\ref{PDC}) has the opposite direction to the actual persistent current. Another distinct feature is that the current avoids armchair junctions of the zigzag nanoring segments. When the states anticross, as the states (iii) and (iv) do in Figs.~\ref{fig2}(a) and \ref{fig2}(b), the PCD becomes a superposition of the currents around the two edges having the shapes as  in Fig.~\ref{fig3}.

Both the electron density distributions and the probability current densities show that the spin-up and spin-down persistent currents have opposite directions at the zigzag edges of the nanoring. Besides, states of the same spin circulate in opposite directions at the inner and outer edges, hence the obtained spectra correspond to helical edge states. Even though time reversal invariance is broken by the magnetic field, a QSH state remains protected for energies below the SO gap ($-\Delta_{SO} \leq E \leq \Delta_{SO}$), which is in agreement with Ref. [\onlinecite{shevtsov12}]. Note that the Brillouin zone is not defined for nanorings, and Chern number $n$ and $\mathbb{Z}_2$ invariant are ill-defined for the edge states. Nonetheless, we can explain their topological origin using the bulk-edge correspondence: the edges of the nanoring separate the topological insulator from the vacuum, which is a trivial insulator with an infinite bandgap, and thus gapless edge states emerge.

\section{Silicene nanorings with armchair edges}\label{IV}

\begin{figure}
	\centering
	\includegraphics[width=15cm]{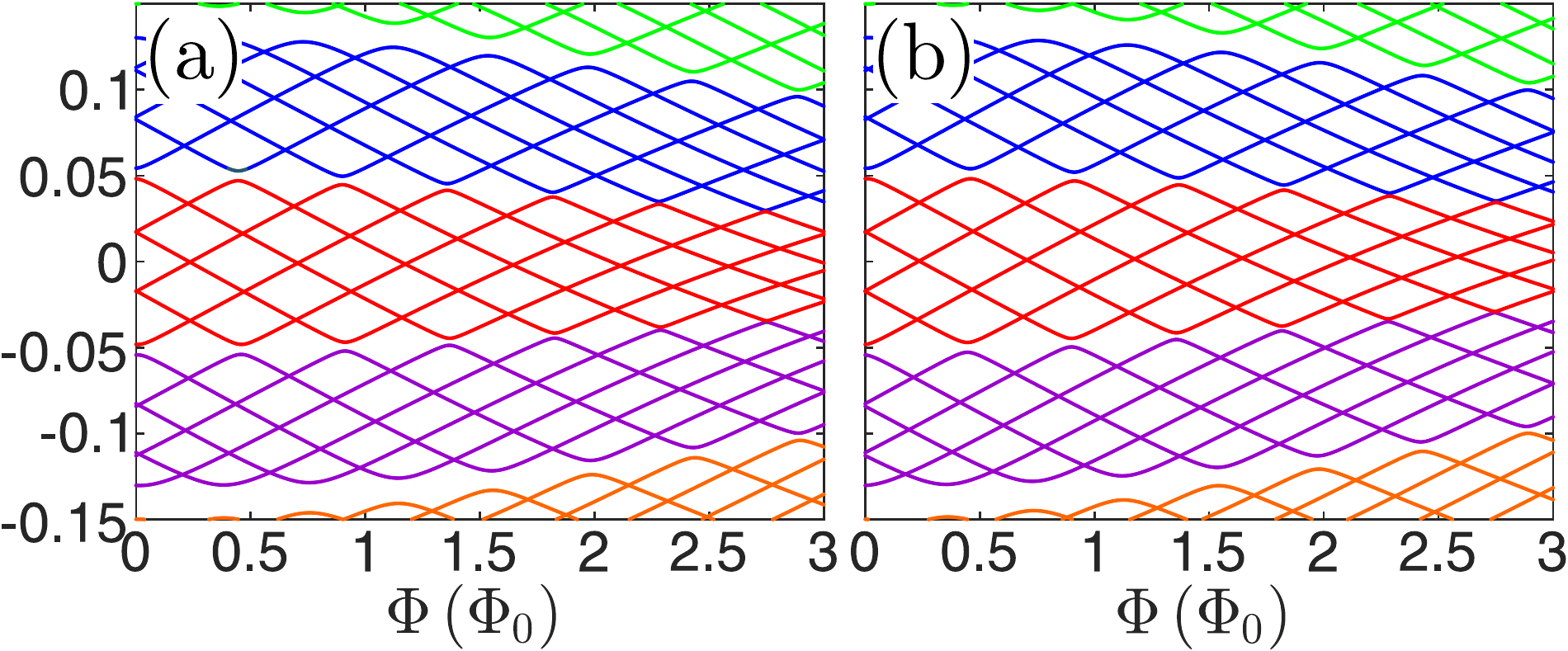}
	\caption{The energy levels of a silicene nanoring with armchair edges as function of magnetic flux for: (a)  spin-up, and (b) spin-down states. Different colors are used to delineate different subbands.}
	\label{fig4}
\end{figure}

The energy spectra of the spin-up and spin-down states as function of magnetic flux for the hexagonal nanoring with armchair edges ($N_E = 35$, $N_I=18$) are shown in Figs. \ref{fig4}(a) and (b), respectively. Opposite to the case of a nanoring with zigzag edges, the shown spectra are similar to the case of a hexagonal graphene nanoring ($\Delta_{SO}=0$) with armchair edges in a perpendicular magnetic field. The organization of the levels is similar as for the zigzag edges, with groups of six levels forming subbands separated from each other by energy gaps. The subband around the Dirac point is rather wide and the particles exhibit an interchanging character from electron-like to hole-like when $B$ varies, as found in wide hexagonal armchair graphene nanorings.\cite{dacosta14} The energy gaps decrease and eventually vanish when $B$ increases as a result of a decreasing magnetic length ($l_B = \sqrt{\hbar/eB}$) leading to a weakening of the coupling between the states localized at the edges responsible for the anticrossings of the energy levels.\cite{bahamon09}

\begin{figure}
	\centering
	\includegraphics[width=10cm]{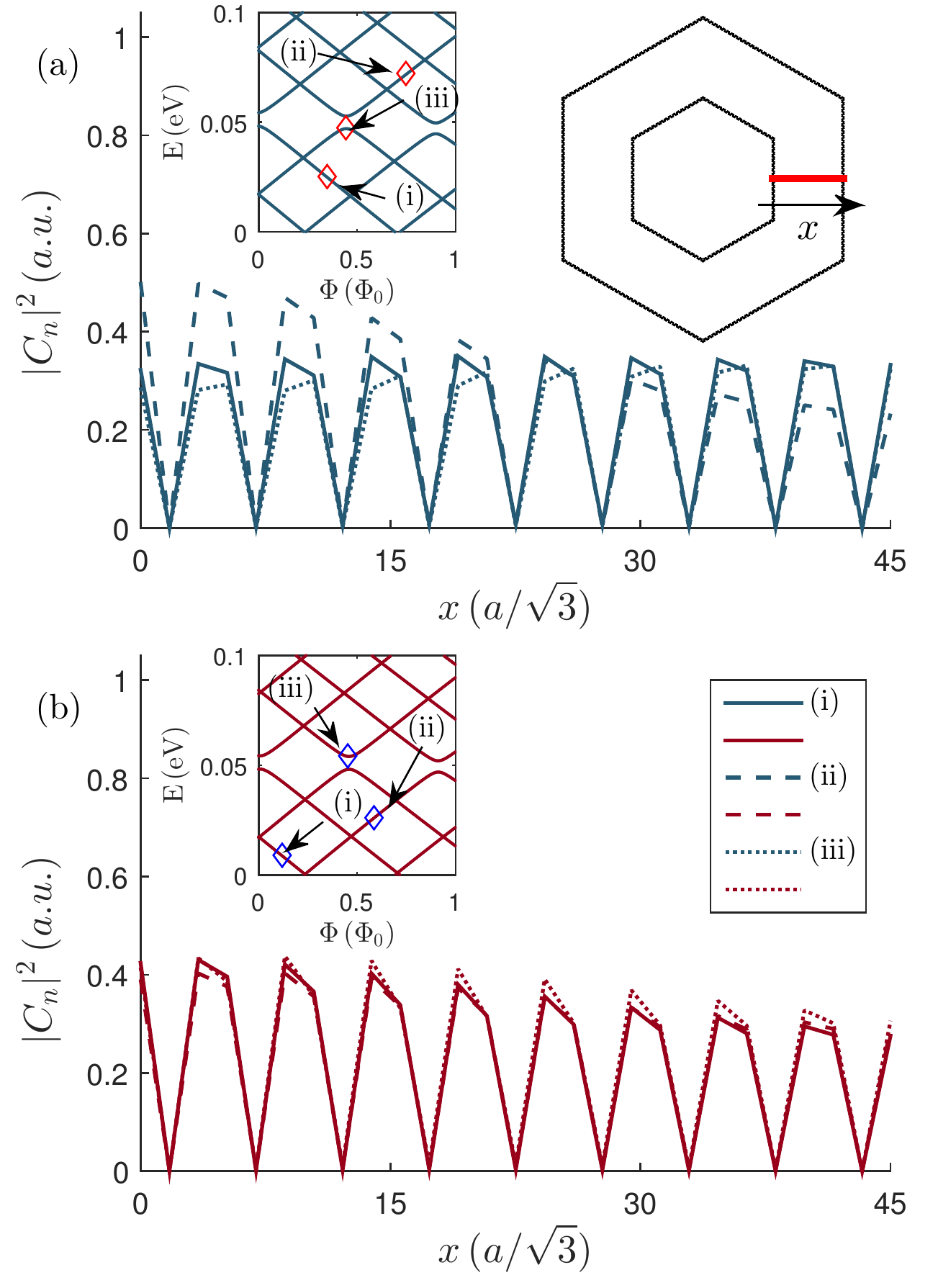}
	\caption{(a) The electron densities of the spin-up states in the armchair ($N_E = 35$, $N_I=18$) nanoring: (i) (solid line), (ii) (dashed line), and (iii) (dotted line). The energies of the depicted states are shown in the left inset, which is a zoom of Fig~\ref{fig4}(a). The red line in the right inset shows the direction along which electron density is plotted for the case of armchair edges. (b) The same as (a), but for the spin-down states.}
	\label{fig5}
\end{figure}

The energies of the spin-up and spin-down states are very close to each other and are almost spin degenerate, even for a large magnetic field. Furthermore, we found that the electrons are almost uniformly distributed throughout the whole ring. Thus we choose to show the electron density along the direction that crosses a ring segment between the ring edges shown by the thick red line in Fig.~\ref{fig5}(a). The electron densities along this direction for a few spin-up and spin-down states are displayed in Figs.~\ref{fig5}(a) and \ref{fig5}(b), respectively. The same types of states are selected as for the zigzag nanorings, and they are indicated by arrows in the insets of Fig.~\ref{fig5}. The electron densities of the state (i), (ii), and (iii) are shown by solid, dashed, and dotted lines, respectively. All three states are evidently similarly localized, with their $|C_n|^2$'s varying almost periodically between the ring edges. This case could be explained by inverse proportionality of the penetration depth of a helical edge channel at armchair edges to the SO gap ($\xi_{\mathrm{arm}}\sim 1/\Delta_{SO}$).\cite{ezawa13, cortes13} It is also related to the disappearance of zero-energy levels of armchair nanoribbons due to the interference between the edge states.\cite{ezawa13} We might conclude that SOC in silicene is not large enough to produce distinctive helical edge-localized states and associated magnetic dipoles in nanorings having armchair edges.

Note that Fig.~\ref{fig4} displays energy levels in the extended range with respect to Fig.~\ref{fig1}, hence it is difficult to compare the two. The energy levels in Fig.~\ref{fig1} are all located in the SO gap, whereas no classification of the states inside and outside the SO gap exist in nanorings with armchair edges. However, one may notice that the average energy of the subband around $E=0$ is almost constant with $B$ in Fig.~\ref{fig4}, whereas average energies of the subbands vary quasilinearly in Fig.~\ref{fig1}, which might be ascribed to the different localization of the states at differently shaped edges. The energy spectrum in nanorings with armchair edges is similar to the energy spectrum of graphene where SOC is vanishingly small. Therefore, for this case we are able to use the same argument as for graphene nanorings, which states that a decrease of magnetic length with increasing magnetic field weakens the coupling between the edges which thus reduces the energy gaps that appear at anticrossings. Furthermore, we inspected how the eigenfunctions vary with the ring width, and we did not find any qualitative difference with respect to the presented results when $N_I$ varies from 19 to 30 for fixed $N_E=35$.

%\begin{figure}[t!]
	%\centering
	%\includegraphics[width=10cm]{figure7_new.eps}
	%\caption{The moduli squared of the eigenfunctions of the spin-up state (i) (solid line), state (ii) (dashed line), and state (iii) (dotted line) along the ring cross section. States are labelled in the inset, which shows a zoom of the spectrum in Fig. \ref{fig6}(a).}
	%\label{fig7}
%\end{figure}

\section{Germanene nanorings with zigzag and armchair edges}\label{V}

\begin{figure}
\centering
\includegraphics[width=15cm]{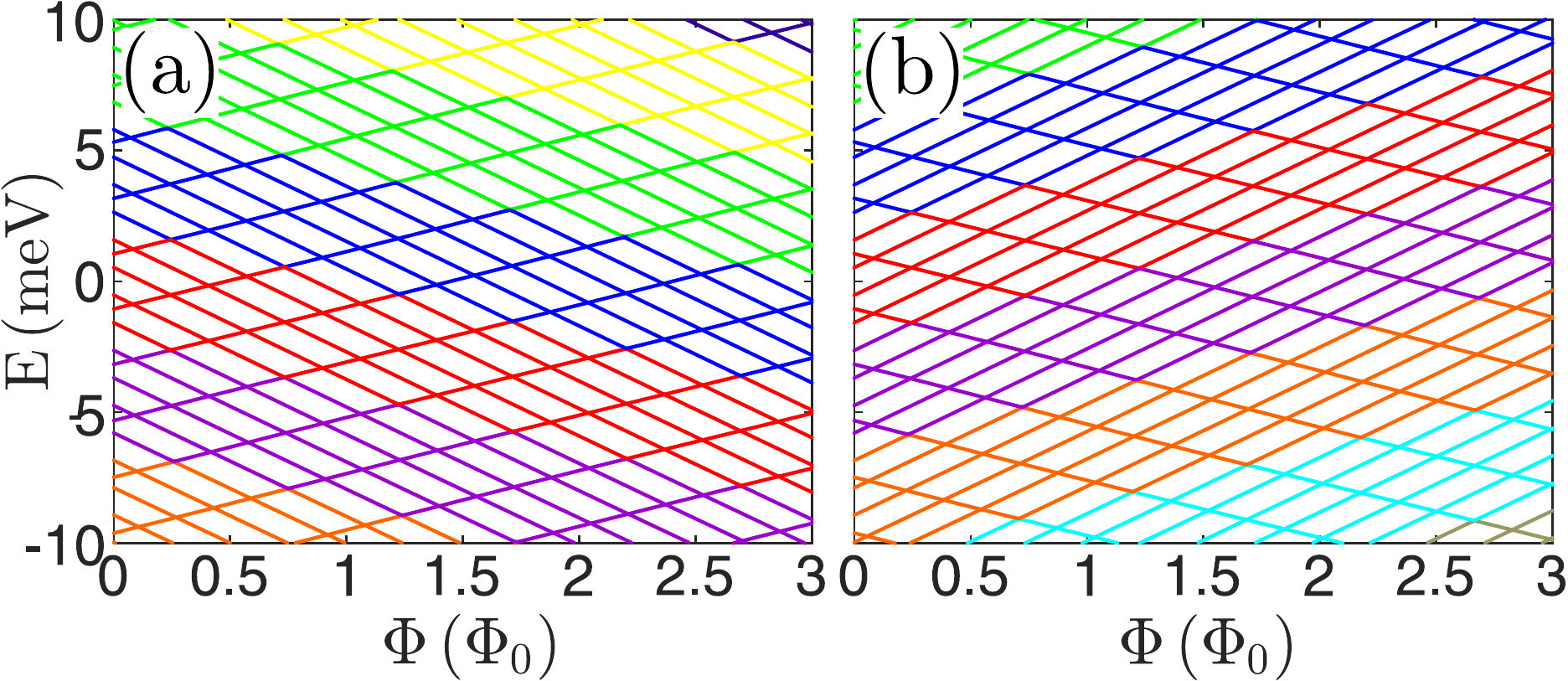}
\caption{The energy levels of a germanene nanoring with zigzag edges as function of magnetic flux for: (a) spin-up, and (b) spin-down states. The subbands are portrayed by using different colors.}
\label{fig6}
\end{figure}

\begin{figure}[t!]
	\centering
	\includegraphics[width=15cm]{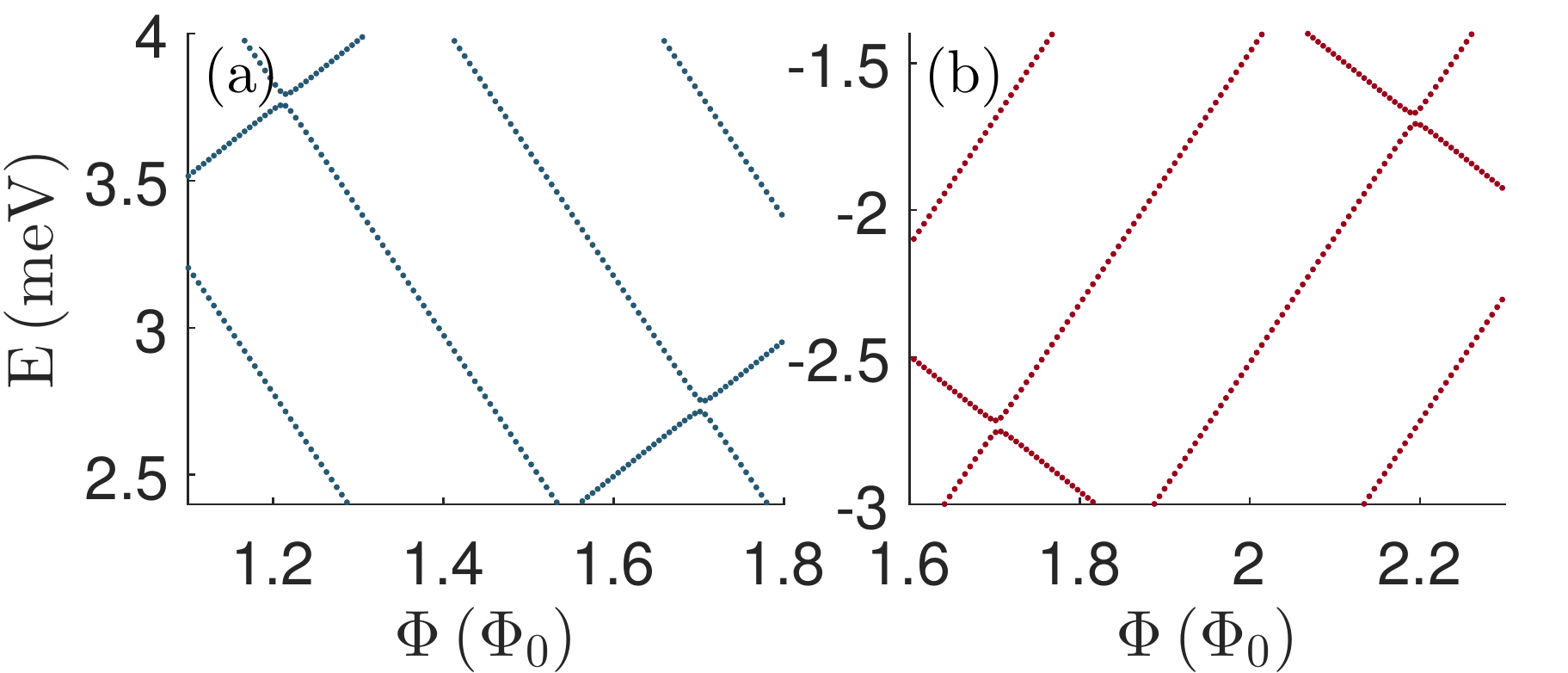}
	\caption{Details of the spectra shown in: (a) Fig.~\ref{fig6}(a) and (b) Fig. \ref{fig6}(b) each exhibiting two sharp anticrossings with extremely small energy gaps.}
	%\label{fig10ab}
	\label{fig7}
\end{figure}

We now extend our analysis to germanene nanorings. These results will be compared with those from silicene rings, with the aim to examine how the magnitude of SOC affects the electronic structure. The energy spectra of the hexagonal germanene nanoring with zigzag edges ($N_E = 60$, $N_I=30$) for the electron spin-up and spin-down states are displayed in Fig. \ref{fig6} as a function of magnetic flux. Here $\Phi = \Phi_0$ corresponds to $B\approx 11$ T. The helical edge spin-up and spin-down states indeed have the anticipated antisymmetry, with the energy levels inside the SO gap decreasing and increasing, respectively. But larger SOC in germanene makes the gaps at anticrossings rather small as zooms of the spectra of Figs. \ref{fig6}(a) and (b) show in Figs. \ref{fig7}(a) and (b), respectively.

\begin{figure}
	\centering
	\includegraphics[width=15cm]{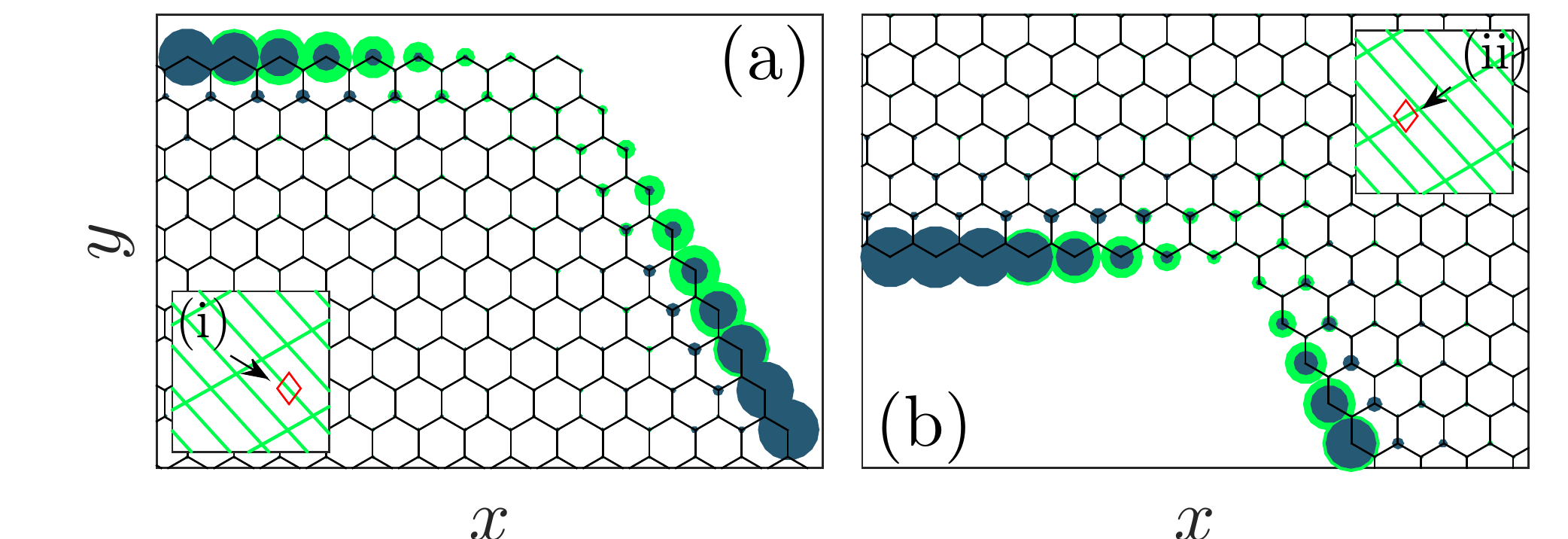}
	\caption{The distributions of the spin-up electron density around the corners: (a) at the outer edge and (b) at the inner edge of the zigzag nanoring. Blue circles centered at the lattice sites represent $|C_n|^2$ for the states in the {\it silicene} nanoring labelled by (i) and (ii) in Fig. \ref{fig2}(a). Green circles centered at the same lattice sites of the zigzag nanoring show $|C_n|^2$ for the spin-up outer and inner edge states in {it germanene} nanoring, which are denoted by (i) and (ii) in the insets.}
	\label{fig8}
\end{figure}

The electron densities for the three states have the same edge localizations as in Fig. \ref{fig2}, and also there exist helical currents at the nanoring edges. The spin-up current $I_{\uparrow}^E$ flows in the counterclockwise direction around the outer edge, whereas the current $I_{\uparrow}^I$ flows in the clockwise direction around the inner edge. The currents of the spin-down electrons $I_{\downarrow}^E$ and $I_{\downarrow}^I$ have directions opposite to the respective spin-up currents. But the regions around the corners are less desolate than in silicence nanorings (see Fig.~\ref{fig8}), i.e. we see that the arms of the density along the edges have higher tendency to establish connections than in silicene nanorings.

%The eigenfunctions of the states (i) and (ii) shown in Figs. \ref{fig10}(a) and \ref{fig10}(b) exhibit the formation of the helical currents at the nanoring edges (Fig. \ref{fig11}). The spin-up current $I_{\uparrow}^E$ flows in the counterclockwise direction around the outer edge (see Fig \ref{fig11}(a)), whereas the current $I_{\uparrow}^I$ has the clockwise direction around the inner edge (see Fig \ref{fig11}(b)). For the spin down, the currents $I_{\downarrow}^E$ and $I_{\downarrow}^I$ have the directions opposite to the respective spin-up currents (see Fig. \ref{fig11}(c) and Fig. \ref{fig11}(d)). By inspecting the shown eigenstates we see that the eigenfunctions in the germanene rings have higher tendency to establish the connections in the rings corners.
%We might consequently infer for even larger SOC, as in the case of stanene, a single continuous distribution might be formed around the edge.

%\begin{figure}[t!]
	%\centering
	%\includegraphics[width=8.6cm]{figure10.eps}
	%\caption{(a) A detail of the spectrum shown in Fig. \ref{fig9}(a). (b) A detail of the spectrum shown in Fig. \ref{fig9}(b). The selected spin-up and spin-down states are indicated by the arrows labelled by (i)-(iv).}
	%\label{fig10ab}
	%\label{fig10}
%\end{figure}

%\begin{figure}
	%\centering
	%\includegraphics[width=8.6cm]{figure11.eps}
	%\caption{The eigenfunctions of the spin-up and spin-down states indicated by the arrows in Fig. \ref{fig10}: (a) the state (i) which forms the spin current $I_{\uparrow}^E$, (b) the state (ii) which forms the spin current $I_{\uparrow}^I$, (c) the state (iii) which forms the spin current $I_{\downarrow}^I$, and (d) the state (iv) which forms the spin current $I_{\downarrow}^E$.}
	%\label{fig11}
%\end{figure}

\begin{figure}
	\centering
	\includegraphics[width=15cm]{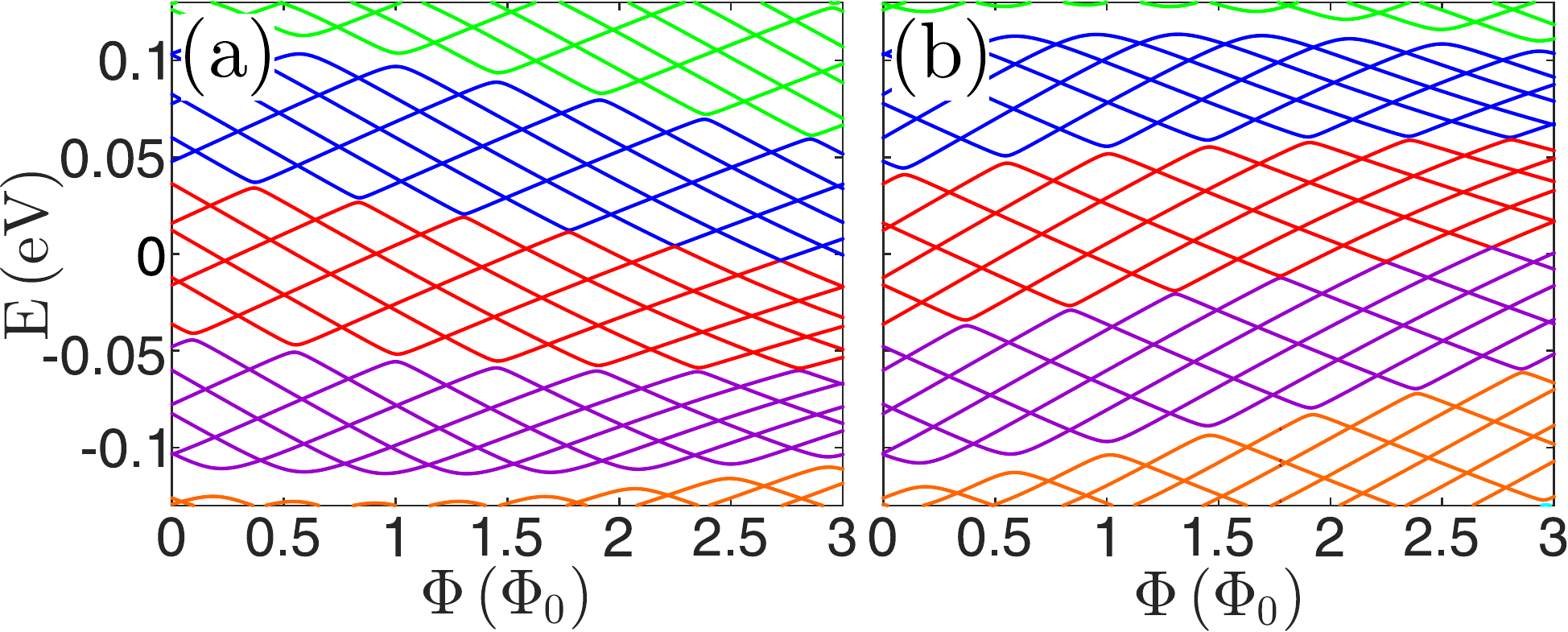}
	\caption{The energy levels of a germanene nanoring with armchair edges as function of magnetic flux for: (a) the spin up and (b) the spin down. Each subband is denoted by a distinct color.}
	\label{fig9}
\end{figure}

\begin{figure}
	\centering
	\includegraphics[width=10cm]{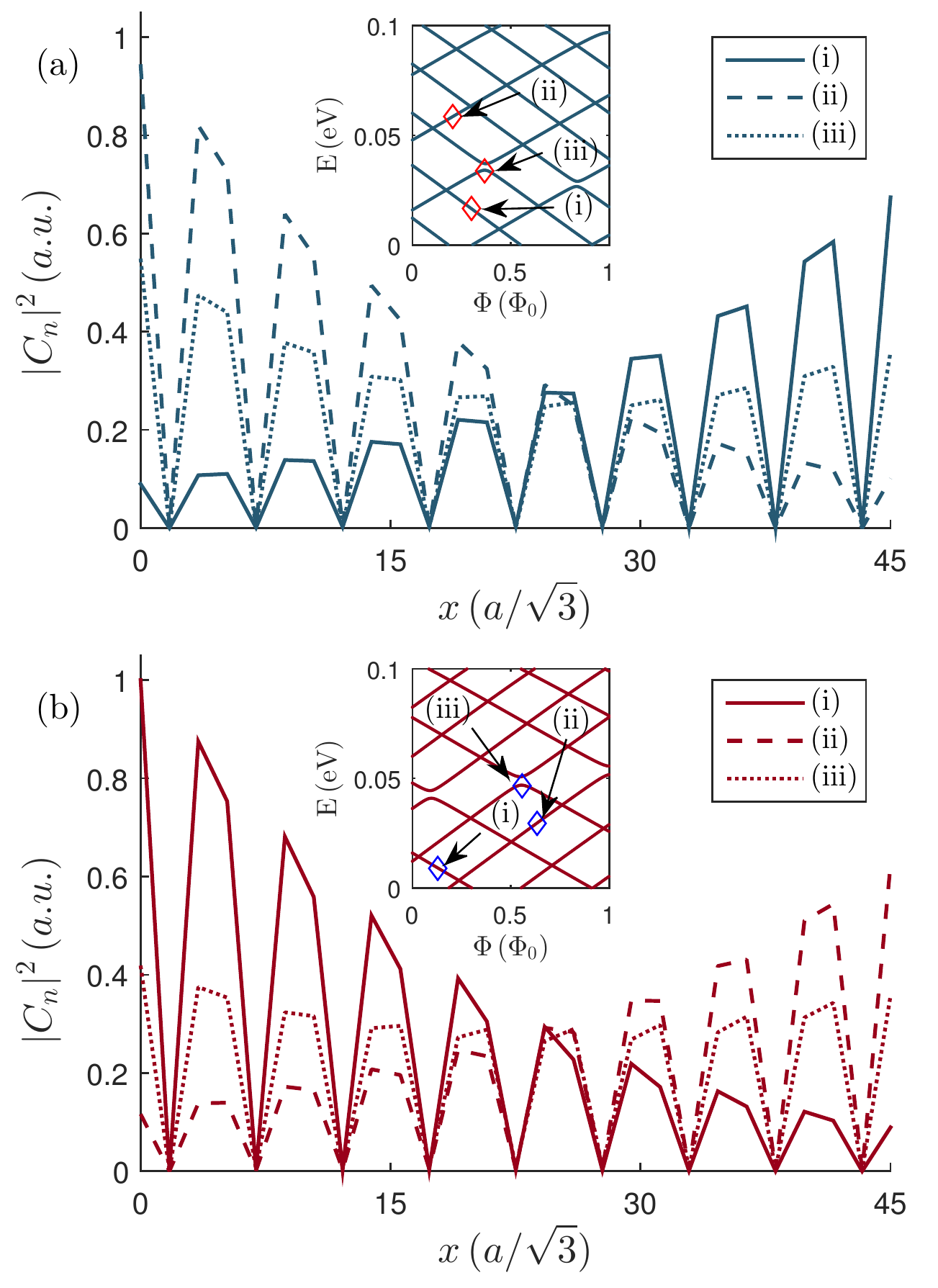}
	\caption{(a) The electron density of the spin-up states: (i) (solid line), (ii) (dashed line), and (iii) (dotted line) between the edges of the ring along the direction shown in the inset of Fig~\ref{fig5}(a). States are labelled in the inset, which shows a zoom of the spectrum in Fig. \ref{fig9}(a). (b) The same as (a), but for the spin-down states.}
	\label{fig10}
\end{figure}

The spectra of the germanene nanoring with armchair edges ($N_E = 35$, $N_I=18$) are organized in subbands of six levels, as in the case of silicene. However, larger SOC in germanene lifts the spin degeneracy even at rather small magnetic field (compare Figs.~\ref{fig4} and \ref{fig9}). It also changes the electron density with respect to the silicene case (see Figs.~\ref{fig10}(a) and \ref{fig10}(b) for the spin up and spin down, respectively), especially of the states (i) and (ii) which are distributed closer to the edges than in the silicene case. It could be related to the state penetration depth which is smaller in germanene. And the states of opposite spins exhibit maximum localization at opposite edges. On the other hand, the anticrossing state (iii) varies almost periodically between the ring edges, similar to the silicene case. Hence, larger spin-orbit interaction in germanene than in silicene brings about substantial changes in the electron localization of the analyzed states in the energy spectrum.

Let us briefly note that we analyzed how computed quantum states in silicene and germanene nanorings having zigzag edges depend on the rings width by varying $N_I$ in the range from 30 to 50 for given $N_E=60$. We were not able to find any regular variations of both the gaps at anticrossings and the subbands widths on the ring dimensions. But the behavior displayed for a few ring dimensions is typical for both silicene and germanene nanorings. Yet, we found that energy gaps at anticrossings are smaller in the germanene than in the silicene case for the same nanoring width.

\section{Conclusion}\label{VI}

We theoretically investigated the electronic structure of hexagonal silicene and germanene nanorings with zigzag and armchair edges in the presence of a perpendicular magnetic field. The electron spectra are calculated as function of magnetic flux through the ring, and also the electron density distributions in the ring planes (for zigzag nanorings) and along the line which crosses one ring segment (for armchair nanorings) are shown and analyzed for both spin-up and spin-down states. The model of a classical magnetic dipole in a magnetic field offers a qualitatively correct description of the helical edge states. The spin-up and the spin-down currents propagate in opposite directions at the inner and outer edges of the zigzag nanorings. In other words, there is a QSH phase at the edges being the boundaries between the nontrivial topological phase and the vacuum. Furthermore, the distributions of the probability current density in the zigzag nanorings show regions near the ring corners where electron transport is avoided. On the other hand, neither silicene nor germanene nanorings with armchair edges exhibit helical edge states. But a larger SOC in germanene removes spin degeneracy in nanorings more efficiently than in their silicene counterparts. As could be inferred from the distributions penetrating the corners of the analyzed zigzag silicene and germanene nanorings, the helical edge states would be more robust in larger nanorings or materials exhibiting larger SOC. The perspective candidate could be stanene, another 2D material that attracted attention recently.\cite{xu2013}

%\begin{figure}
	%\centering
	%\includegraphics[width=11cm]{AC_crosssection.eps}
	%\caption{Nanoring with armchair edges ($N_E = 35$, $N_I=18$). Red line indicates the cross-section of an arm along which eigenfucntions shown in Figs \ref{fig7},\ref{fig8},\ref{fig13}, and \ref{fig14} are plotted.}
	%\label{fig15}
%\end{figure}

%Moreover,  the armchair edges exhibit the presence of helical edge states.

%The Kane-Mele model for finite honeycomb lattice is employed and influence of magnetic field is taken into account through the %Peierls phase.

\begin{acknowledgments}This work was supported by Erasmus+ and the Serbian Ministry of Education, Science and Technological Development (Project No. III45003).
\end{acknowledgments}

\end{document}